\documentclass[onecolumn,12pt]{article} 

\usepackage[margin=1.15in]{geometry}
  
 \usepackage{setspace}
 \usepackage{float} 
\usepackage{graphicx}
\usepackage{amsmath}
\usepackage{amssymb}
\usepackage{dsfont}
\usepackage[english]{babel}
\usepackage[latin1]{inputenc}
\usepackage[T1]{fontenc}

\usepackage{algorithm,algorithmic}

\usepackage{caption,subcaption,float} 

\usepackage[authoryear, round]{natbib}

 \usepackage[
                  breaklinks = true,
                 colorlinks = true,
                 linkcolor = red,
                 urlcolor  = blue,
                 citecolor = blue,
                 anchorcolor = green,
                 ]{hyperref}

\graphicspath{
{Results/}
{Results/BSSR/}
{Results/BMSSR/}
}
\newcommand{\bc}{\mathbf{c}}

\newcommand{\bB}{\mathbf{B}}
\newcommand{\bO}{\mathbf{0}}

\newcommand{\bb}{\mathbf{b}}
\newcommand{\be}{\mathbf{e}}

\newcommand{\bR}{\mathbf{R}}

\newcommand{\bI}{\mathbf{I}}

\newcommand{\bS}{\mathbf{S}}

\newcommand{\bT}{\mathbf{T}}

\newcommand{\bV}{\mathbf{V}}

\newcommand{\bx}{\mathbf{x}}
\newcommand{\bX}{\mathbf{X}}
\newcommand{\by}{\mathbf{y}}
\newcommand{\bY}{\mathbf{Y}}
\newcommand{\bz}{\mathbf{z}}

\newcommand{\bspi}{\boldsymbol{\pi}}
\newcommand{\bsmu}{\boldsymbol{\mu}}

\newcommand{\bsalpha}{\boldsymbol{\alpha}}
\newcommand{\bsbeta}{\boldsymbol{\beta}}

\newcommand{\bsSigma}{\boldsymbol{\Sigma}}
\newcommand{\bssigma}{\boldsymbol{\sigma}}

\newcommand{\bsvPsi}{\boldsymbol{\varPsi}}

\newcommand{\bsxi}{\boldsymbol{\xi}}

\newcommand{\Identity}{\textbf{I}}

\newcommand{\Pro}{\mathbb{P}}

\newcommand{\N}{\mathcal{N}}

\date{}

\begin{document}

\title{Bayesian mixtures of spatial spline regressions}
\author{Faicel Chamroukhi}
 \maketitle
\begin{center}
Aix Marseille Universit\'e, CNRS, ENSAM, LSIS, UMR 7296, 13397 Marseille, France\\
Universit\'e de Toulon, CNRS, LSIS, UMR 7296, 83957 La Garde, France\\
\href{mailto:chamroukhi@univ-tln.fr}{chamroukhi@univ-tln.fr}
\end{center}
 

\begin{abstract} 
This work relates the framework of model-based clustering for spatial functional data where the data are surfaces. We first introduce a Bayesian spatial spline regression model with mixed-effects (BSSR) for modeling spatial function data. The BSSR model is based on Nodal basis functions for spatial regression and accommodates both common mean behavior for the data through a fixed-effects part, and variability inter-individuals thanks to a random-effects part. Then, in order to model populations of spatial functional data issued from heterogeneous groups, we integrate the BSSR model into a mixture framework. The resulting model is a Bayesian mixture of spatial spline regressions with mixed-effects (BMSSR) used for density estimation and model-based surface clustering. The models, through their Bayesian formulation, allow to integrate possible prior knowledge on the data structure and constitute a good alternative to recent mixture of spatial spline regressions model estimated in a maximum likelihood framework via the expectation-maximization (EM) algorithm. The Bayesian model inference is performed by Markov Chain Monte Carlo (MCMC) sampling. We derive two Gibbs sampler to infer the BSSR and the BMSSR models and apply them on simulated surfaces and a real problem of handwritten digit recognition using the MNIST data set. The obtained results highlight the potential benefit of the proposed Bayesian approaches for modeling surfaces possibly dispersed in particular in clusters. 
\end{abstract}

\noindent {\bf key-words}:
Bayesian spatial spline regression; Bayesian mixture of spatial spline regression; Surface approximation; Model-based surface clustering; Gibbs sampling; Spatial functional data analysis; Handwritten digit recognition.

\section{Introduction}

Functional data analysis (FDA) \citep{Ramsay2005,ramsayandsilvermanAppliedFDA2002,FerratyANDVieuBook} is the paradigm of data analysis in which the individuals are functions (e.g., curves or surfaces) rather than vectors of reduced dimension. 
Most of the classical analyses directly consider the data to be analyzed as vectors. 
However, in many areas of application, including signal and image processing, functional imaging, handwritten text recognition, genomics, diagnosis of
complex systems, etc., the analyzed data are often available in the form of (discretized) values of functions or curves (e.g., times series, waveforms, etc) and surfaces (2D-images, spatio-temporal data, etc) which makes them very structured. This ``functional" aspect of the data adds additional difficulties in the the analysis compared to the case of a classical multivariate analysis. It is fortunately possible to overcome these difficulties encountered in multivariate (non functional) analysis techniques, by formulating ``functional'' models that explicitly integrate the functional form of the data, rather than directly considering
them as vectors. 
This is the FDA framework for data clustering, classification and regression.
The key tenet of FDA is to treat the data not just as multivariate observations but as (discretized) values of smooth functions. This approach allows to more fully exploit the structure of the data. 
In this framework, several models have been introduced to model univariate and multivariate functional data for clustering or  classification.
Among these models, one distinguishes the finite mixture model-based ones, on which we focus in this paper. 
Indeed, the flexibility, easy interpretation and efficiency of finite mixture models \citep{McLachlan2000,SylviaFruhwirthBook2006,TitteringtonBookMixtures} 
in multivariate analysis, has lead to a growing investigation for adapting them to the
framework of FDA. 
For example, one can cite the following papers, among many others, which relate probabilistic generative models for FDA  
\citep{Devijver2014,
Jacques2014,
chamroukhi_fmda_neucomp2013,
Delaigle2012,
Bouveyron2011,
Same2011,
chamroukhi_et_al_neurocomputing2010,
chamroukhi_PhD_2010,
chamroukhi_et_al_NN2009,
Liu2009,
Gaffney2004,
Gaffneythesis,
garetjamesJASA2003,
garetjamesANDtrevorhastieJRSS2001}.

These models have however mainly focused on the study of univariate or multivariate functions. For the case of spatial functional data, 
\cite{Malfait2003,Ramsay2011,Sangalli2013,Nguyen2014} proposed methods to deal with surfaces.
In particular, the recent  approach proposed by \cite{Nguyen2014} for clustering and classification of surfaces  is based on the
regression spatial spline regression as in \cite{Sangalli2013} in a mixture of linear mixed-effects model framework as in \citep{CeleuxMLMM2005}. 
\cite{Nguyen2014} indeed extended the functional data analysis framework for univariate functions to the analysis of spatial functions (i.e. surfaces) by introducing a spatial spline regression (SSR) model and a mixture of spatial spline regressions (MSSR) model, to respectively model homogeneous surfaces and heterogeneous surfaces with a clustering structure.
The SSR model with mixed-effects is tailored to spatial regression data with both fixed-effects and random-effects. 
The mixture of spatial spline regression (MSSR) is dedicated to surface clustering, as in \citep{garetjamesJASA2003} for curve clustering, while the mixture of spatial spline regression discriminant analysis (MSSR-DA) is deditcated to curve discrimination, in a similar way as the discriminant analysis approach for curves proposed by \cite{garetjamesANDtrevorhastieJRSS2001}.
%
The usual used tool for model estimation is  maximum  likelihood estimation (MLE) by using the expectation-maximization (EM) algorithm \citep{McLachlanEM2008,dlr}.
While MLE via the EM algorithm is the standard way to fit  finite mixture-based models, a common alternative is the Bayesian inference, that is, the maximum a posteriori (MAP) estimation  by using in general Markov Chain Monte Carlo (MCMC) sampling.

Indeed, the Bayesian inference framework has also led to intensive research in the filed of mixture models and Bayesian methods for mixtures have become popular due to advances in both methodology and computing power.
The application of Bayesian methods to mixture models are included namely in \cite{Robert1994}, and \cite{AndrewGelman2003}. 
Some key papers on the Bayesian analysis of mixtures are \cite{DieboltAndRobert1994}, \cite{EscobarANDWest_95_BayesianMixtures} and \cite{RichardsonANDGreen97} and \cite{CeleuxJASA2000}.
One can cite for example the following references among many others that deal with Bayesian mixture modeling and inference: 
\citep{Robert1994,
Stephens_thesis_97,
Bensmail_model_based_clust97,
Ormonenti_IEEENN_98,
Stephens98bayesiananalysis_mixtures,
bayes_modeling_inference_mixtures2005,
SylviaFruhwirthBook2006,
fraley_and_raftery_2007}

While the MLE approaches maximizes the model likelihood, the Bayesian (MAP) approach maximizes adds a prior distribution over the model parameters and then maximizes the posterior parameter distribution. 
The MAP estimation can still be performed by the EM algorithm (namely in the case of conjugate priors) as in \cite{fraley_and_raftery_2007} or by MCMC sampling, such as the Gibbs sampler \cite{Neal1993,rafterygibbs,rafterygibbs,Bensmail_model_based_clust97,bayes_modeling_inference_mixtures2005,Robert2011}.  
For the Bayesian analysis of regression data, \cite{LenkANDDeSarbo2000} introduced a Bayesian inference for finite mixtures of generalized linear models with random effects. Int their mixture model, each component is a regression model with a random-effects parts and the model is dedicated to multivariate regression data. 
  
In this paper, we present a probabilistic Bayesian formulation to model spatial functional data by extending the approaches of \cite{Nguyen2014} and apply the proposal to surface approximation and clustering. The model  is also related to the random-effects mixture model of \cite{LenkANDDeSarbo2000} in which we explicitly add mixed-effects  and derive it for spatial functional data by using the Nodal basis functions (NBFs). 
The NBFs \citep{Malfait2003} used in \cite{Ramsay2011,Sangalli2013,Nguyen2014}  represent an extension of the univariate B-spline bases to bivariate surfaces. 
We thus introduce  the Bayesian spatial spline regression with mixed-effects (BSSR) for fitting a population of homogeneous surfaces and the Bayesian mixtures of SSR (BMSSR) for fitting populations of heterogeneous surfaces organized in groups. The BSSR model is first applied in surface approximation. Then, the BMSSR model is applied in model-based surface clustering by considering the real-world handwritten digits from the MNIST data set \citep{LecunMnist}.

This paper is organized as follows. 
Section \ref{sec: Related work} provides a description of recent related work  on mixture of spatial spline regressions. Then, in Section \ref{sec: BSSR}, we present the BSSR model and its inference technique using Gibbs sampling. Then, in Section 
\ref{sec: BMSSR}, we present the Bayesian mixture formulation, that is, the BMSSR model, and show how to apply it in model-based clustering of surfaces. A Gibbs sampler is  derived to estimate the BMSSR model parameters. 
In section \ref{sec: application}, we apply  the proposed models on simulated surfaces and on a real handwritten digit recognition problem.  Finally, in Section \ref{sec: conclusion}, we draw some conclusions 
 and mention some future possible directions for this research.
 
\section{Mixtures of spatial spline regressions with mixed-effects} 
\label{sec: Related work}
This section is dedicated to related work on mixture of spatial spline regressions  (SSR)   with mixed-effects (MSSR), introduced by \cite{Ng_and_McLachlan_2014}.
We first describe the regression model with linear mixed-effects and its mixture formulation, in the general case, and then describe the models for spatial regression data. 

\subsection{Regression  with mixed-effects}

The miexd-effects regression models  (see for example \cite{LairdAndWare1982},
 \cite{VerbekeAndLesaffre1996} and \cite{XuAndHedeker2001}),  are appropriate when the standard regression model (with fixed-effects) can not sufficiently explain the data. For example, when representing dependent data arising from related individuals or when data are gathered over time on the same individuals. In that case, the mixed-effects regression model is more appropriate as it includes both fixed-effects and  random-effects terms. In the linear mixed-effects regression model, 
 the $m_i \times 1$ response $\by_i = (y_{i1},\ldots,y_{im_i})^T$ is modeled as:
\begin{equation}
\by_i = \bX_i \bsbeta + \bT_i \bb_i+ \be_i
\label{eq: mixed-effects regression}
\end{equation}where the $p \times 1$ vector $\bsbeta$ is the usual unknown fixed-effects regression coefficients vector describing the population mean, 
$\bb_i$ is a $q\times 1$ vector of unknown subject-specific regression coefficients corresponding to individual effects, independently and identically distributed (i.i.d) according to the normal distribution $\N(\bsmu_i,\bR_i)$ and independent from the  $m_i \times 1$ error terms $\be_i$ which are distributed according to $\N(\bO,\bsSigma_i)$, and
$\bX_i$ and $\bT_i$ are respectively $m_i \times p$ and $m_i \times q$ known covariate matrices.  
A common choice for the noise covariance-matrix is to take a diagonal matrix $\bsSigma_i = \sigma^2\Identity_{m_i}$ where $\Identity_{m_i}$ denotes the $m_i \times m_i$ identity matrix.
%
%
Thus, under this model, the joint distribution of  the observations $\by_i$ and the random effects $\bb_i$  is the following joint multivariate normal distribution (see for example \cite{XuAndHedeker2001}):
\begin{eqnarray}
\left[\begin{array}{c}
\by_i \\ 
\bb_i
\end{array} \right] \sim
\N\left( 
\left[\begin{array}{c}
\bX_i \bsbeta + \bT_i \bsmu_i \\ 
\bsmu_i
\end{array} \right],
\left[\begin{array}{cc}
\sigma^2 \Identity_{m_i} + \bT_i \bR_{i} \bT_i^T & \bT_i \bR_{i}\\ 
\bR_{i} \bX_i^T & \bR_{i} 
\end{array} \right]
\right).
\label{eq: joint of y and b mixed-effects}
\end{eqnarray}Then, from (\ref{eq: joint of y and b mixed-effects}) it follows that the observations $\by_i$  are marginally distributed according to the following normal distribution 
(see \cite{VerbekeAndLesaffre1996} and \cite{XuAndHedeker2001}):
\begin{eqnarray}
f(\by_i|\bX_i,\bT_i;\bsvPsi) = \N (\by_{i};\bX_i \bsbeta + \bT_i \bsmu_i, \sigma^2\Identity_{m_i} + \bT_i \bR_i \bT_i^T).
\label{eq: marginal of y mixed-effects regression}
\end{eqnarray}

\subsection{Mixture of regressions with mixed-effects} 

The regression model with mixed-effects (\ref{eq: mixed-effects regression}) can be integrated into a finite mixture framework to deal with regression data arising from a finite number of groups. 
The resulting mixture of regressions model with linear mixed-effects \citep{VerbekeAndLesaffre1996,XuAndHedeker2001,CeleuxMLMM2005,NgEtAll2006}  is   a mixture model where every component $k$ ($k=1,\ldots, K$) is a regression model with mixed-effects given by (\ref{eq: mixed-effects regression}), $K$ being the number of mixture components. Thus, the observation $\by_i$ conditionally on each component $k$ is modeled as:
\begin{equation}
\by_{i} = \bX_i \bsbeta_k + \bT_i \bb_{ik} + \be_{ik}
\label{eq: mixture of regressions with mixed-effects}
\end{equation}
where $\bsbeta_k$, $\bb_{ik}$ and $be_ {ik}$ are respectively the the 
the fixed-effects regression coefficients,
the random-effects regression coefficients for individual $i$,
 and the error terms, for component $k$. 
The random-effect coefficients $\bb_{ik}$ 
are i.i.d according to  $\N(\bsmu_{ki},\bR_{ki})$ and are independent from the error terms $\be_{ik}$ which follow the distribution $\N(\bO,\sigma_k^2\Identity_{m_i})$. 
%
%
Let $Z_i$ denotes the categorical random variable representing the component memebership for the $i$th observation. 
Thus,  conditional on the component $Z_i =k$, the observation $\by_i$ and the random effects $\bb_i$ have the following joint multivariate normal distribution:
\begin{eqnarray}
\left[\begin{array}{c}
\by_i \\ 
\bb_i
\end{array} \right]\Bigg|_{Z_i =k} \sim
\N\left( 
\left[\begin{array}{c}
\bX_i \bsbeta + \bT_i \bsmu_k \\ 
\bsmu_k 
\end{array} \right],
\left[\begin{array}{cc}
\sigma_{k}^2 \Identity_{m_i} + \bT_i \bR_{ki} \bT_i^T & \bT_i \bR_{ki}\\ 
\bR_{ki} \bX_i^T & \bR_{ki} 
\end{array} \right]
\right)
 \label{eq: joint of y and b mixture of mixed-effects}
\end{eqnarray}and thus the observation $\by_i$  are marginally distributed according to the following normal distribution 
(see \cite{VerbekeAndLesaffre1996} and \cite{XuAndHedeker2001}): 
\begin{eqnarray}
f(\by_i|\bX_i,\bT_i, Z_i = k;\bsvPsi_k) = \N (\by_{i};\bX_i \bsbeta_k + \bT_i \bsmu_{ki},  \bT_i \bR_{ki} \bT_i^T + \sigma_k^2\Identity_{m_i}).
\label{eq: marginal of y mixed-effects regression}
\end{eqnarray}The unknown parameter vector of this component-specific density is given by:
$$\bsvPsi_k=(\bsbeta^T_{k}, \sigma_{k}^2, \bsmu^T_{k1},\ldots,\bsmu^T_{kn}, \text{vech}(\bR_{k1})^T,\ldots, \text{vech}(\bR_{kn})^T)^T$$  where vech is the half-vectorization operator which produces the lower triangular portion of the symmetric matrix it operates on. 
Thus, the marginal distribution of $\by_i$ unconditional on component memberships is given
by the following mixture distribution:
\begin{equation}
f(\by_i|\bX_i,\bT_i;\bsvPsi) = \sum_{k=1}^K \pi_k \, \N (\by_{i};\bX_i \bsbeta_k + \bT_i \bsmu_{ki},  \bT_i \bR_{ki} \bT_i^T + \sigma_k^2\Identity_{m_i}) 
\label{eq: density of mixture of regressions with mixed-effets (MRMM)}
\end{equation}where the $\pi_k$'s given by $\pi_k = \Pro(Z_i = k)$ for $k=1,\ldots,K$ represent the mixing proportions which are non-negative and sum to 1.
The unknown mixture model parameters   given by the parameter vector 
$$\bsvPsi = (\pi_1,\ldots,\pi_{K-1},\bsvPsi^T_1,\ldots,\bsvPsi^T_K)^T$$  where 
$\bsvPsi_k$ is the parameter vector of component $k$, are usually estimated, given an i.i.d sample of $n$ observations, by maximizing the  observed-data log-likelihood
\begin{equation}
\log L(\bsvPsi) =  \sum_{i=1}^n  \log  \sum_{k=1}^K \pi_k \, \N (\by_{i};\bX_i \bsbeta_k + \bT_i \bsmu_{ki},  \bT_i \bR_{ki} \bT_i^T + \sigma_k^2\Identity_{m_i}) 
\label{eq: log-lik MRMM}
\end{equation}via the expectation-maximization (EM) algorithm \citep{dlr,McLachlanEM2008,VerbekeAndLesaffre1996,XuAndHedeker2001,CeleuxMLMM2005,NgEtAll2006}.

\subsection{Mixtures of spatial spline regressions with mixed-effects}
 

For spatial regression data, \cite{Nguyen2014} introduced the spatial spline regression with liner mixed-effects (SSR). The model is given by (\ref{eq: mixed-effects regression}) where the covariate matrices, which are assumed to be identical in \cite{Nguyen2014}, that is, $\bT_i=\bX_i$ and denoted by $\bS_i$, in this spatial case, represent a spatial structure are calculated from the Nodal Basis Function (NBF) \citep{Malfait2003}. Note that in what follows we will denote the number of columns of $\bS_i$ by $d$. 
The NBF idea is an extension of the  B-spline bases  used in general for univariate or multivariate functions,  to bivariate surfaces and was first introduced by \cite{Malfait2003} and then used namely in \cite{Ramsay2011} and \cite{Sangalli2013} for surfaces. 

In \cite{Nguyen2014}, it is assumed that the random-effects are centered with isotropic covariance matrix common to all the individuals, that is $\bb_i \sim \N(\bO,\xi^{2}\bI_{m_i})$.  
Thus, from (\ref{eq: marginal of y mixed-effects regression}) it follows that under the spatial spline regression model with linear mixed-effects, the density of the observation $\by_i$ is given by
\begin{eqnarray}
f(\by_i|\bS_i;\bsvPsi) = \N (\by_{i};\bS_i \bsbeta, \xi^{2}\bS_i\bS_i^T + \sigma^2\Identity_{m_i}).
\label{eq: marginal of y spatial mixed-effects regression}
\end{eqnarray}It follows that under the mixture of spatial spline regression models with linear mixed-effects, the density of $\by_i$ is given by:
\begin{equation}
f(\by_i|\bS_i;\bsvPsi) = \sum_{k=1}^K \pi_k \, \N (\by_{i};\bS_i \bsbeta_k,  \xi^2_{k} \bS_i \bS_i^T + \sigma_k^2\Identity_{m_i})
\label{eq: density of mixture of spatial regressions with mixed-effets (MSRMM)}
\end{equation}where the model parameter vector is given by:
$$\bsvPsi = (\pi_1,\ldots,\pi_{K-1},\bsbeta^T_{1},\ldots,\bsbeta^T_{K},\sigma_{1}^2, \ldots,\sigma_{K}^2,\xi_{1}^2,\ldots,\xi_{K}^2)^T.$$  
Both of models are fitted by using the EM algorithm. In particular, for the mixture of spatial spline regressions, the EM algorithm maximizes the following observed-data log-likelihood:
\begin{equation}
\log L(\bsvPsi)= \sum_{i=1}^n \log \sum_{k=1}^K \pi_k \, \N (\by_{i};\bS_i \bsbeta_k,  \xi^2_{k} \bS_i \bS_i^T + \sigma_k^2\Identity_{m_i}).
\label{eq: log-lik MSRMM}
\end{equation}
More details on the EM developments  for the two models can be found in detail in \cite{Nguyen2014}. Note that \cite{Nguyen2014} assumed a common noise variance $\sigma^2$ for all the mixture components in (\ref{eq: density of mixture of spatial regressions with mixed-effets (MSRMM)}) and hence  in (\ref{eq: log-lik MSRMM}).

\section{Bayesian spatial spline regression with   mixed-effects (BSSR)}
\label{sec: BSSR} 
We introduce a  Bayesian probabilistic  approach to the spatial spline regression model with mixed-effects presented in \cite{Nguyen2014} in a maximum likelihood context.  
The proposed model is thus the Bayesian spatial spline regression with linear mixed-effects (BSSR) model.   
We first present the model, the parameter distributions and then derive the Gibbs sampler for parameter estimation.
 
\subsection{The model}

The Bayesian spatial spline regression with mixed-effects (BSSR) model   is defined by: 
\begin{equation}
\by_i = \bS_i (\bsbeta + \bb_i) + \be_i
\label{eq: Bayesian mixed-effects regression}
\end{equation}where the model parameters in this Bayesian framework are assumed to be random variables with specified prior distributions, and the spatial covariates matrix $\bS_i$ is computed from the Nodal basis functions. 
We first describe the Nodal basis functions and then continue the model formulation derivation.

\begin{figure}[H]
\centering
\includegraphics[width=8cm]{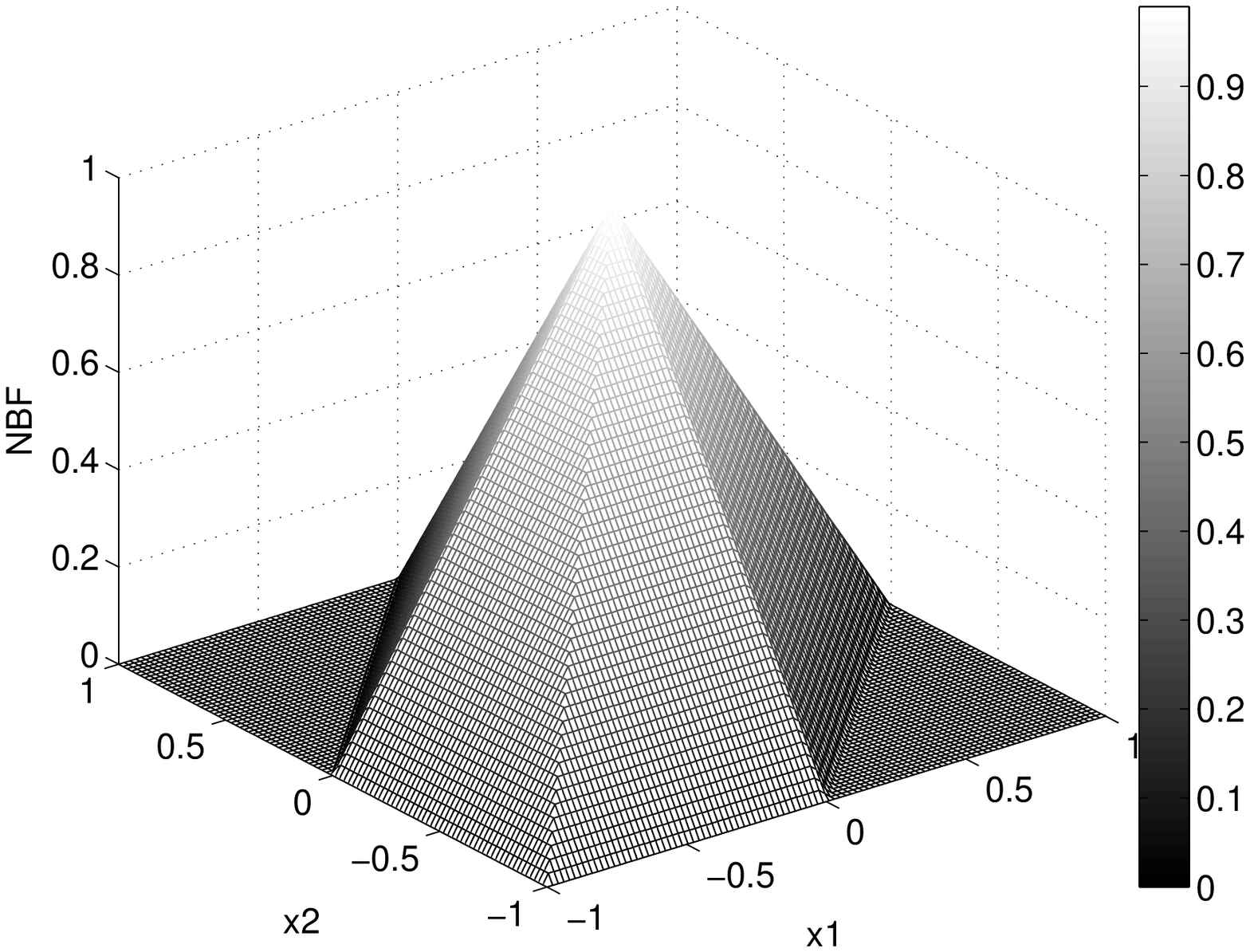}
\caption[]{Nodal basis function $s(\bx,\bc,\delta_{1},\delta_{2})$, where $\bc=(0,0)$ and $\delta_1=\delta_2 =1$.}
\label{fig:NBF example}
\end{figure}

Introduced by \cite{Malfait2003}, the idea of Nodal basis functions (NBFs) extends  the use of B-splines for univariate function approximation \citep{Ramsay2005}, to the approximation of surfaces.
For a fixed number of basis functions $d$, defined on a regular grid with regularly spaced points $ c(l)$ ($l=1,\ldots,d$) of the domain we are working on,  with $d$ defined as $ d=d_1 d_2 $ where $d_1$ and $d_2$ are respectively the columns and rows number of nodes, the $i$th surface can be approximated using piecewise linear Lagrangian triangular finite element 
NBFs constructed as (e.g see \cite{Sangalli2013,Nguyen2014}): 
\begin{scriptsize} 
\begin{equation}
s(\bx,\bc,\delta_{1},\delta_{2}) =
\begin{cases}

-\dfrac{x_{2}}{\delta_{2}}+\dfrac{c_{2}+\delta_{2}}{\delta_{2}}
& \text{if } \bx \in \left\lbrace (x_1,x_2): c_1 < x1 \leq c_1 + \delta_1, \dfrac{\delta_2}{\delta_1} x_1 + \dfrac{\delta_1c_2-\delta_2 c_1} {\delta_1} \leq x_2 < c_2 + \delta_2\right\rbrace \\

-\dfrac{x_{1}}{\delta_{1}}+\dfrac{c_{1}+\delta_{1}}{\delta_{1}}
& \text{if } \bx \in \left\lbrace (x_1,x_2): c_1 < x1 \leq c_1 + \delta_1, c_2 \leq x_2 <   \dfrac{\delta_2}{\delta_1} x_1 + \dfrac{\delta_1 c_2-\delta_2 c_1} {\delta_1}\right\rbrace\\

-\dfrac{x_{1}}{\delta_{1}}+ \dfrac{x_{2}}{\delta_{2}}+
\dfrac{\delta_1 \delta_2+\delta_2 c_1 -\delta_1 c_2}{\delta_1 \delta_2}
& \text{if } \bx \in \left\lbrace (x_1,x_2): c_1 < x1 \leq c_1 + \delta_1,\dfrac{\delta_2}{\delta_1} x_1 + \dfrac{\delta_1 c_2 - \delta_2 c_1 -\delta_1 \delta_2}{\delta_1} \leq x_2 <c_2 \right\rbrace\\

\dfrac{x_{2}}{\delta_{2}} + \dfrac{\delta_{2} - c_{2}}{\delta_{2}} 
& \text{if } \bx \in \left\lbrace  (x_1,x_2): c_1 - \delta_1 \leq x_1 < c_1,
c_2-\delta_2 \leq x_2 \leq \dfrac{\delta_2}{\delta_1} x_1 + \dfrac{\delta_1 c_2 -\delta_2 c_1}{\delta1} \right\rbrace \\ 

\dfrac{x_{1}}{\delta_{1}} + \dfrac{\delta_{1} - c_{1}}{\delta_{1}} 
& \text{if } \bx \in \left\lbrace (x_1,x_2): c_1 - \delta_1 \leq x_1 < c_1,
\dfrac{\delta_2}{\delta_1} x_1 + \dfrac{\delta_1 c_2 - \delta_2 c_1 }{\delta_1} < x_2 \leq c_2 \right\rbrace\\

\dfrac{x_{1}}{\delta_{1}} -  \dfrac{x_{2}}{\delta_{2}}+
\dfrac{\delta_1 \delta_2+\delta_1 c_2 - \delta_2 c_1}{\delta_1 \delta_2}
& \text{if } \bx \in \left\lbrace (x_1,x_2):c_1 - \delta_1 \leq x_1 < c_1, 
c_2 < x_2 \leq \dfrac{\delta_2}{\delta_1 } x_1 + \dfrac{\delta_1 c_2 + \delta_1 \delta_2 - \delta_2 c_1}{\delta_1} \right\rbrace\\

0 & \text{otherwise}

\end{cases}
\end{equation}
\end{scriptsize}where $\bx_{ij}=(x_{ij1},x_{ij2})$ are the two spatial coordinates of $y_{ij}$, $\bc=(c_1,c_2)$ denotes a node center parameter and $\delta_{1} $ and $\delta_{1} $ are respectively the vertical and horizontal  shape parameters representing the distances between two consecutive centers.   
Thus, this construction leads to the following $m_i \times d$ spatial covariates matrix:
\begin{equation}
\bS_{i}=
\begin{pmatrix}
s({\bx_1;\bc_1}) & s({\bx_1;\bc_2}) & \cdots &s({\bx_1;\bc_d}) \\
s({\bx_2;\bc_1})& s({\bx_2;\bc_2})& \cdots &s({\bx_2;\bc_d}) \\
\vdots  & \vdots  & \ddots & \vdots  \\
s({\bx_{m_i};\bc_1}) & s({\bx_{m_i};\bc_2}) & \cdots & s({\bx_{m_i};\bc_d})
\end{pmatrix}
\label{eq:matrixNBF}
\end{equation}
where $ s(\bx;\bc)  $ is a shortened notation of the NBF $s(\bx,\bc,\delta_{1},\delta_{2})$ (the shape parameters $\delta_1$ and $\delta_2$ being constant).
An example of a NBF function defined on the rectangular domain $(x_1,x_2) \in [-1,1]\times [-1,1]$ with a single node $\bc=(0,0)$ and $\delta_1=\delta_2 =1$ is presented in the Figure~\ref{fig:NBF example}. 

\begin{figure}[H]
\centering
\includegraphics[width=5cm]{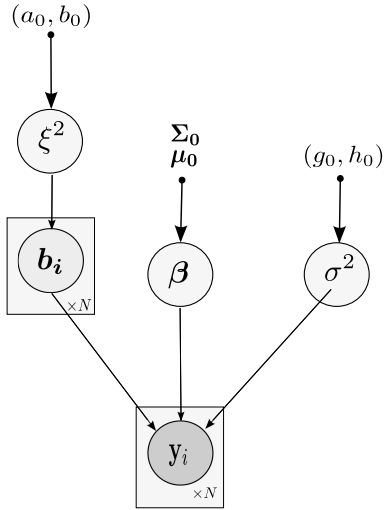}
\caption[Graphical  representation of the BMSSR model]{Graphical  representation of the proposed BMSSR model}
\label{fig:BMSSRgraphicalmodel}
\end{figure}

The model parameters of the proposed Bayesian model, which are given by the parameter vector $\bsvPsi = (\bsbeta^T,\sigma^2,\bb_1,\ldots,\bb_n,\xi^2)^T$ are assumed to be unknown random variables with the following prior distributions. 
We use conjugate priors for ease as those mostly used priors in the literature for example as in \citep{DieboltAndRobert1994}\cite{RichardsonANDGreen97} \citep{Stephens98bayesiananalysis_mixtures}. The used priors for the parameters are as follows:
%
\begin{equation}
\begin{tabular}{lll}
$\bsbeta$ & $\sim$ & $\N(\bsmu_{0},\bsSigma_{0})$ \\
$\bb_{i}|\xi^{2}$ &$\sim$ & $\N(\boldsymbol{0}_{d},\xi^{2}\bI_{d})$\\
$\xi^{2}$ &$\sim$ & $IG(a_{0},b_{0})$\\
$\sigma^{2}$ &$\sim$ & $IG(g_{0},h_{0})$
\end{tabular} 
\label{eq: BSSR prior}
\end{equation}

where $(\bsmu_{0},\bsSigma_{0}) $ are the hyper-parameters of the normal prior over the fixed-effects coefficients, $ \xi^{2} $ is the variance of the normal distribution over the random-effect coefficients,  $a_{0}$ and $b_{0}$ (respectively $g_{0}$ and $h_{0}$) are respectively the shape and scale parameters of the Inverse Gamma $(IG)$ prior over the variance $\xi^{2}$ (respectively $\sigma^{2}$).
Figure~\ref{fig:BMSSRgraphicalmodel} shows the graphical representation of the proposed BSSR model for a set of homogeneous functions $(\by_1,\ldots,\by_n)$. 

We use MCMC sampling for the Bayesian inference of the model. MCMC sampling is indeed one the most commonly used inference techniques in Bayesian analysis of mixtures, in particular the Gibbs sampler (e.g see \cite{DieboltAndRobert1994}). 
\subsection{Bayesian inference using Gibbs sampling}
\label{ssec: Bayesian inference BSSR}

In order to implement the Gibbs sampler, we first derive the full conditional posterior distributions of the model parameters.
Due to the chosen conjugate hierarchical prior (\ref{eq: BSSR prior}) presented in the previous section, the full conditional posterior distributions can then be found analytically as shown in detail in the \href{http://chamroukhi.univ-tln.fr/appendix-bayes-SSRM.pdf}{Appendix A}.
The full conditional distribution of each of the model parameter are given in the following subsections. 
We use the notation $ |... $   to denote a conditioning of the parameter in question on all the other parameters and the observed data. 

\subsubsection{Full conditional distribution of the fixed-effects coefficient vector $\bsbeta$}
Applying the Bayes theorem to the joint distribution leads to the following posterior over the fixed-effects regression coefficients $\bsbeta$:
$p(\bsbeta|...)= p(\bY|\bsbeta,\bB,\sigma^{2}) p(\bsbeta)$. 
Thus, the $\bsbeta$'s posterior distribution is given by the following normal distribution:
\begin{equation}
\bsbeta|... \sim \N(\boldsymbol{\nu}_{0},\bV_{0})
\label{eq: BMSSR posterior of beta}
\end{equation}
with 
\begin{eqnarray*}
\bV_{0}^{-1}&=&\bsSigma_{0}^{-1}+ \frac{1}{\sigma^{2}}\sum_{i=1}^{n}\bS_{i}^{T}\bS_{i},\\
\boldsymbol{\nu}_{0}&=& \bV_{0}\left(\frac{1}{\sigma^{2}}\sum_{i=1}^{n}\left(\by_{i}-\bS_{i}\bb_{i}\right)-\bsSigma_{0}^{-1}\boldsymbol{\mu_{0}}\right).
\end{eqnarray*} 

\subsubsection{Full conditional distribution of the random-effect coefficient vector $\bb_i$}
By using the same reasoning as for the fixed-effects regression coefficients,  the  posterior of the random-effects coefficients is calculated as:
$p(\bb_{i}|...)= p(\by_i|\bsbeta,\bb_i,\sigma^{2})p(\bb_{i}|\xi^{2})$
and is thus given by the following normal distribution:
\begin{equation}
\bb_ {i}|... \sim \N(\boldsymbol{\nu}_{1},\bV_{1})
\label{eq: BMSSR posterior of bi}
\end{equation}
with:  
\begin{eqnarray*}
\bV_{1}^{-1}&=& \frac{1}{\sigma^{2}}\bS_{i}^{T}\bS_{i}+\frac{1}{\xi^{2}},\\
\boldsymbol{\nu}_{1}&=& \bV_{1}\left(\frac{1}{\sigma^{2}}\bS_{i}^{T}(\by_{i}-\bS_{i}\bsbeta)\right).
\end{eqnarray*}

\subsubsection{Full conditional distribution of the noise variance $\sigma^2$}
For the noise variance $\sigma^2$ which has an inverse Gamma prior,  the posterior given by
$p(\sigma^{2}|...) = p(\bY|\bsbeta, \bB,\sigma^{2})p(\sigma^{2})$ is the following inverse Gamma distribution: 
\begin{equation}
{\sigma^{2}}|... \sim IG(g_{1},h_{1})
\label{eq: BMSSR posterior of sigma}
\end{equation} 
with   
\begin{eqnarray*}
g_{1} &=& g_{0}+\frac{n}{2}, \\
h_{1} &=& h_{0}+\frac{\sum_{i=1}^{n} \left( \by_{i}- \bS_{i}\bsbeta-\bS_{i}\bb_{i} \right) ^{T}\left( \by_{i}- \bS_{i}\bsbeta-\bS_{i}\bb_{i}\right)}{2}\cdot
\end{eqnarray*} 

\subsubsection{Full conditional distribution of the random-effect variance $\xi^{2}$}
The same reasoning is used to derive the  posterior of the random-effect variance $\xi^{2}$. The posterior distribution for the parameter is given by $p(\xi^{2}|...) \propto p(\bB|\xi^{2})p(\xi^{2})$ which leads to the following posterior inverse Gamma distribution:
 \begin{equation}
{\xi^{2}}|... \sim IG \left( a_{1}, b_{1} \right) 
\label{eq: BMSSR posterior of xi}
\end{equation} 
with: 
\begin{eqnarray*}
a_{1} & = & a_{0}+\frac{n}{2}, \\
b_{1} & = & b_{0} + \frac{\sum_{i=1}^{n}\bb_{i}^{T}\bb_{i}}{2}\cdot
\end{eqnarray*}
Algorithm~\ref{algo: Gibbs for BSSR} summarizes the implementation of the Gibbs sampler for the proposed BMSSR model. Each  sample of the Gibbs sampler is drawn from the above posterior distributions. 
\begin{algorithm}[h]
\caption{Gibbs sampling for the Bayesian spatial spline regression model with mixed-effects (BMSSR)}
\label{algo: Gibbs for BSSR}
\begin{algorithmic}
\STATE {\bfseries Inputs:} The  observations $\bY=(\by_{1},\ldots,\by_{n})$ and the spatial spline regression matrices $(\bS_1,\ldots,\bS_n)$
\STATE {\bfseries Initialize:} \\ 
fix the model hyper parameters: $(\bsmu_{0},\bsSigma_{0},g_{0},h_{0},a_{0},b_{0})$ \\ 
\STATE   initialize the model parameters: $(\bsbeta^{(0)},\bB^{(0)}, {\xi^{2}}^{(0)}, {\sigma^{2}}^{(0)})$

\FOR{$t=1$ {\bfseries to} $\#$Gibbs samples}
\STATE 1.  Sample the random-effects variance: ${\xi^{2}}^{(t)} \sim IG \left( a_{0}+\frac{n}{2},  b_{0} + \frac{\sum_{i=1}^{n} {\bb_{i}^{(t-1)}}^{T}\bb_{i}^{(t-1)}}{2}\right)$

\STATE 2. Sample  the noise variance\\ ${\sigma^2}^{(t)} \sim IG \left(g_{0}+\frac{n}{2} , h_{0}+\frac{\sum_{i=1}^{n} \left( \by_{i}- \bS_{i}\bsbeta^{(t-1)}-\bS_{i}\bb_i^{(t-1)} \right)^{T}\left( \by_{i}- \bS_{i}\bsbeta^{(t-1)}-\bS_{i}\bb_i^{(t-1)}\right)}{2}\right) $

\STATE 3. Sample the fixed-effects coefficients vector $\beta^{(t)}\sim \N(\boldsymbol{\nu}^{(t)}_{0},\bV^{(t)}_{0}) $ with\\
${\bV_{0}^{-1}}^{(t)}=\bsSigma_{0}^{-1}+\frac{1}{{\sigma^{2}}^{(t)}}\sum_{i=1}^{n}\bS_{i}^{T}\bS_{i}$, \\
$\boldsymbol{\nu}^{(t)}_{0}= \bV_{0}\left(\frac{1}{{\sigma^{2}}^{(t)}}\sum_{i=1}^{n}\bS_{i}^{T} ( \by_{i}-\bS_{i}\bb_{i}^{(t-1)})-\bsSigma_{0}^{-1}\boldsymbol{\mu_{0}}\right)$ 

\FOR{$i=1$ {\bfseries to} $n$}
\STATE 4. Sample the random-effects coefficients vector $\bb_{i}^{(t)}\sim \N(\boldsymbol{\nu}^{(t)}_{1},\bV^{(t)}_{1})$ with \\
${\bV_{1}^{-1}}^{(t)}= \frac{1}{{\sigma^2}^{(t)}}\bS_{i}^{T}\bS_{i}+\frac{1}{{\xi^2}^{(t)}}$\\
$\boldsymbol{\nu}^{(t)}_{1}= \bV_{1}\left(\frac{1}{{\sigma^2}^{(t)}}\bS_{i}^{T}(\by_{i}-\bS_{i}\bsbeta^{(t)})\right)$

\ENDFOR

\ENDFOR
\end{algorithmic}
\end{algorithm}

\section{Bayesian mixture spatial spline regressions with mixed-effects (BMSSR)}
\label{sec: BMSSR}

The BMSSR model presented previously is dedicated to learn from a single or a set of homogeneous spatial functional data. 
However, when the data present a natural grouping aspect, this may be restrictive, and its extension to accommodate clustered data is needed.
We therefore integrate the BMSSR model into a mixture framework. This is mainly motivated by a clustering prospective. The resulting model is therefore a Bayesian mixture of spatial spline regression with mixed-effects (BMSSR) and is described in the following section.
\subsection{The model}
Consider that there are $K$ sub-populations within the data set $\bY=(\by_1,\ldots,\by_n)$. The proposed BMSSR model has the following stochastic representation. Conditional on component $k$, the individual $\by_i$ is modeled by a BSSR model as:
\begin{equation}\label{eq:BMMSSR}
\by_{i}= \bS_{i} (\bsbeta_{k}+ \bb_{ik})+\be_{ik}.        
\end{equation}
%
%
%
Thus, a $K$ component Bayesian mixture of spatial spline regression models with mixed-effects (BMSSR) has the following density:
\begin{equation}
f (\by_{i} |\bS_{i};\bsvPsi ) = \sum_{k=1}^{K}\pi_{k}~\N\left(\by_{i}; \bS_{i}(\bsbeta_{k}+\bb_{ik}),\sigma^{2}_{k}\bI_{m_{i}} \right)
\label{eq: density BMSSR}
\end{equation}
where the parameter vector of the model is given by 
$$\bsvPsi = (\pi_1,\ldots,\pi_{K-1},\bsbeta^T_{1},\ldots,\bsbeta^T_{K},\bB^T_{1},\ldots,\bB^T_{K},\sigma_{1}^2, \ldots,\sigma_{K}^2,\xi_{1}^2,\ldots,\xi_{K}^2)^T, $$
$\bB_k = (\bb^T_{1k},\ldots,\bb^T_{nk})^T$ being the vector of the random-effect coefficients of the $k$th BSSR component. 

\begin{figure}[h]
\centering
\includegraphics[width=5cm]{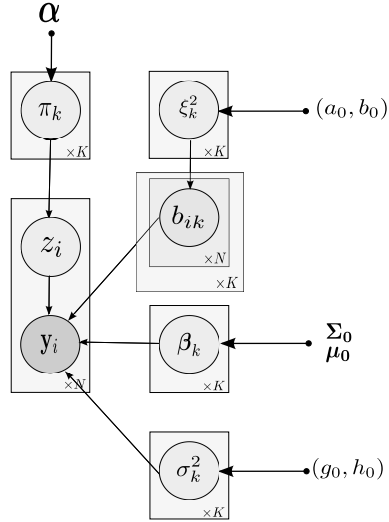}
\caption{Graphical representation of the proposed BMSSR model.}
\label{fig:BMMSSRgraphicalmodel}
\end{figure}

The BMSSR model is indeed composed of BSSR components, each of them has parameters 
$\bsvPsi_k = (\bsbeta^T_{k},\bB^T_{k},\sigma_{k}^2, \xi_{k}^2)^T$
and a mixing proportion parameter $\pi_k$.
Therefore, conditional on component $k$, the parameter priors are defined as in the BSSR model presented (\ref{eq: BSSR prior}) in the previous section. For the BMSSR model, we therefore just need to specify the distribution on the mixing proportions $\bspi = (\pi_1,\ldots,\pi_K)$ which follow the Multinomial distribution in the generative model of the non-Bayesian mixture. 
We use a conjugate prior as for the other parameters, thats is, a Dirichlet prior with hyper-parameters $\bsalpha =(\alpha_{1},\ldots,\alpha_{K})$. The hierarchical prior from for the BMSSR model parameters is therefore given by:
\begin{equation}
\begin{tabular}{lll}
$\pi$ & $\sim$ & $Dir(\alpha_{1},\ldots,\alpha_{K})$ \\
$\bsbeta_{k}$ & $\sim$ & $\N(\bsbeta_{k}|\boldsymbol{\mu_{0}},\Sigma_{0})$\\
$\bb_{ik}|\xi_{k}^{2}$ & $\sim$ & $\N(\bb_{ik}|\boldsymbol{0}_{d},\xi_{k}^{2}\bI_{d})$\\
$\xi_{k}^{2}$ &$\sim$ & $IG(\xi_{k}^{2}|a_{0},b_{0})$\\
$\sigma_{k}^{2}$ & $\sim$ & $IG(\sigma_{k}^{2}|g_{0},h_{0})$.
\end{tabular} 
\label{eq: BMSSR prior}
\end{equation} 
Figure~\ref{fig:BMMSSRgraphicalmodel}  shows the graphical representation of the proposed BMSSR model for a set of heterogeneous functions $(\by_1,\ldots,\by_n)$. 


\subsection{Bayesian inference using Gibbs sampling}
\label{ssec: Bayesian inference BMSSR}
In this section we derive the full conditional posterior distributions needed for the Gibbs sampler to infer the model parameters. Further mathematical calculation details for these posterior distributions are given in \href{http://chamroukhi.univ-tln.fr/appendix-bayes-SSRM.pdf}{Appendix B}.
Consider the vector of augmented parameters, which is the vector of parameters 
$(\bspi^T, \bsbeta^T,\bB^T, {\bssigma^2}^T,{\bsxi^2}^T)^T$ where $\bspi = (\pi_{1},\ldots,\pi_{K})^T$, $\bsbeta=(\bsbeta_{1}^T,\ldots,\bsbeta_{K}^T)^T$, $\bssigma^2 = (\sigma^{2}_{1},\ldots,\sigma^{2}_{K})^T$,  and $\bsxi^2 = (\xi^{2}_{1},\ldots,\xi^{2}_{K})^T$,
augmented by the unknown components labels $\bz = (z_1,\ldots,z_n)$ and the data $\bY$. 
Let us also introduce the binary latent component-indicators $z_{ik}$ 
such that $z_{ik}=1$ iff $z_i = k$, $z_i$ being the hidden label of the mixture component from which the $i$th observation is generated. 
Similarly to the case of Bayesian multivariate Gaussian mixtures,
the posterior distributions of the allocation variables $\bz$ and the mixing proportions $\bspi$ are
Multinomial and Dirichlet, and are as follows
(see for example  \cite[Section 6.4]{RobertBayesianChoiceBook}. 

\subsubsection{Full conditional distributions of the discrete indicator variables $\bz$}

The posterior distributions of the allocation variables $\bz$is given by the following Multinomial distribution with parameters the posterior probabilities of the component labels $z_i$, that is:
\begin{equation}
Z_i|... \sim \text{Mult}(1;\tau_{i1},\ldots,\tau_{iK})
\end{equation}with $\tau_{ik}$ ($1\leq k \leq K$) the posterior probability that the $i$th observation is issued from mixture component $k$:
 \begin{equation}
\tau_{ik} = \Pro(Z_i = k |\by_i, \bS_i;\bsvPsi) = \frac{\pi_{k}~\N\left(\by_{i}|\bS_{i}(\bsbeta_{k}+ \bb_{ik}),\sigma^{2}_{k}\bI_{m_{i}} \right)}
{\sum_{l=1}^K\pi_{l}~\N\left(\by_{i}|\bS_{i}(\bsbeta_{l}+ \bb_{il}),\sigma^{2}_{l}\bI_{m_{i}} \right)}\cdot
\label{eq: postprob BMSSR}
\end{equation}

\subsubsection{Full conditional distribution of the mixing proportions $\bspi$}
The mixture proportions, which have a Dirichlet prior of parameter $\bsalpha$, have the following Dirichlet posterior distribution:
\begin{eqnarray}
\bspi|...
&\sim & \text{Dir}\left(\alpha_{1}+n_1,\ldots, \alpha_{K}+n_K\right)
\label{eq: posterior of pi BMSSR}
\end{eqnarray}with $n_{k}=\sum_{i=1}^{n}z_{ik}$ being the number of observations originated from component $k$.

\subsubsection{Full conditional distribution of the fixed-effects coefficient vectors $\bsbeta_{k}$}
The full conditional posterior distribution of the fixed-effects coefficient vector $\bsbeta_k$, which has normal prior distribution, is obtained By applying the Bayes theorem to the joint distribution and leads to the following normal posterior distribution $p(\bsbeta_{k}|...) = p(\bY|\bsbeta_{k},\bb_{k},\sigma^{2}_{k},\bz) p(\bsbeta_{k})$ which is specified as:
\begin{eqnarray}
\bsbeta_ {k}|... &\sim& \N(\boldsymbol{\nu}_{0},\bV_{0})
\label{eq: posterior of betak BMSSR}
\end{eqnarray}
where
\begin{eqnarray}
\bV_{0}^{-1}&=&\bsSigma_{0}^{-1}+\frac{1}{\sigma_{k}^{2}}\sum_{i=1}^{n}z_{ik}~\bS_{i}^{T}\bS_{i}, \nonumber \\
\boldsymbol{\nu}_{0}&=& \bV_{0}\left(\frac{1}{\sigma_{k}^{2}}\sum_{i=1}^{n}z_{ik}~\bS_{i}^{T}\left( \by_{i}-\bS_{i}\bb_{ik}\right)-\bsSigma_{0}^{-1}\boldsymbol{\mu_{0}}\right). 
\end{eqnarray}

\subsubsection{Full conditional distribution of the random-effects coefficient vectors $\bb_{ik}$} 
The posterior distribution over the random-effects coefficients $\bb_{ik}$ is computed similarly and is given by the following posterior normal distribution thanks to the conjugate normal prior:
$p(\bb_{ik}|...) =  p(\by_i|\bsbeta_{k},\bb_{ik},\sigma^{2}_{k},z_i=k)p(\bb_{ik}|\xi_{k}^{2})$, that is: 
\begin{eqnarray}
\bb_ {ik}|... &\sim& \N(\boldsymbol{\nu}_{1},\bV_{1})
\label{eq: posterior of bik BMSSR}
\end{eqnarray}
where
\begin{eqnarray*}
\bV_{1}^{-1}&=& \frac{1}{\sigma^{2}_{k}}\bS_{i}^{T}\bS_{i}+\frac{1}{\xi_{k}^{2}},\\
\boldsymbol{\nu}_{1}&=& \bV_{1}\left(\frac{1}{\sigma^{2}_{k}}\bS_{i}^{T}(\by_{i}-\bS_{i}\bsbeta_{k})
\right).
\end{eqnarray*}

\subsubsection{Full conditional distribution of the noise variances $\sigma^2_{k}$}
The Inverse Gamma prior on $\bssigma^2$ leads to the following posterior $p(\sigma_{k}^{2}|...) = p(\bY|\bsbeta, \bb_{k},\sigma_{k}^{2},\bz)p(\sigma_{k}^{2})$, which is also an Inverse Gamma distribution given by:
\begin{eqnarray}
{\sigma^{2}_{k}}|... &\sim& IG(g_{1},h_{1})
\label{eq: posterior of sigmak BMSSR}
\end{eqnarray} 
with:
\begin{eqnarray}
g_{1} &=& g_{0}+\frac{1}{2}\sum_{i=1}^{n} z_{ik}, \nonumber\\
h_{1} &=& h_{0}+\frac{\sum_{i=1}^{n}{z_{ik}} \left( \by_{i}- \bS_{i}\bsbeta_{k}-\bS_{i}\bb_{ik} \right) ^{T}\left( \by_{i}- \bS_{i}\bsbeta_{k}-\bS_{i}\bb_{ik}\right)}{2}\cdot \nonumber
\end{eqnarray}

\subsubsection{Full conditional distribution of the random-effects variances $\xi_{k}^{2}$}
The same reasoning is used to derive the  posterior of the random-effect variances $\xi_{k}^{2}$, for which the prior is an Inverse Gamma. The posterior  is in this case calculated as $p(\xi_{k}^{2}|...)= p(\bb_{k}|\xi_{k}^{2})p(\xi_{k}^{2})$ and is given by the following Inverse Gamma distribution:
\begin{eqnarray}
{\xi^{2}_{k}}|... &\sim& IG \left( a_{1}, b_{1} \right) 
\label{eq: posterior of xik BMSSR}
\end{eqnarray}
with 
\begin{eqnarray}
a_{1} & = & a_{0}+\frac{n}{2},\nonumber \\
b_{1} & = & b_{0} + \frac{\sum_{i=1}^{n} \bb_{ik}^{T}\bb_{ik} }{2}\cdot \nonumber
\end{eqnarray}

The pseudo-code \ref{algo: algo: Gibbs for BMSSR} summarizes the Gibbs sampler to infer the parameters of the proposed Bayesian mixture of spatial spline regressions with mixed-effects (BMSSR). Each Gibbs sample is drawn from the above posterior distributions. 
\begin{algorithm}[H]
\caption{Bayesian mixture of spatial spline regressions with mixed-effects (BMSSR).}
\label{algo: algo: Gibbs for BMSSR}
\begin{algorithmic}
\STATE {\bfseries Inputs:} The observations $\bY=(\by_{1},\ldots,\by_{n})$ and the spatial spline regression matrices $(\bS_1,\ldots,\bS_n)$ and the number of mixture components $K$
\STATE {\bfseries Initialize:} \\
the model hyper parameters: $(\bsalpha, \bsmu_{0},\bsSigma_{0},g_{0},h_{0},a_{0},b_{0})$ \\
\STATE   the model parameters: $(\bspi,\bsbeta,\bB, \bsxi^{2}, \bssigma^{2})$
\FOR{$t=1$ {\bfseries to} $\#$Gibbs samples}
\FOR{$i=1$ {\bfseries to} $n$}
\STATE 1. Sample the allocation variables: $z^{(t)}_i \sim \text{Mult}(1;\tau_{i1}^{(t)},\ldots,\tau_{iK}^{(t)})$ with the posterior probabilities $\tau_{ik}^{(t)}$ calculated according to 
$\tau^{(t)}_{ik} = \frac{\pi^{(t)}_{k}~\N\left(\by_{i}|\bS_{i}(\bsbeta^{(t)}_{k}+ \bb^{(t)}_{ik}),{\sigma^{2}}^{(t)}_{k}\bI_{m_{i}} \right)}
{\sum_{l=1}^K\pi^{(t)}_{l}~\N\left(\by_{i}|\bS_{i}(\bsbeta^{(t)}_{l}+ \bb^{(t)}_{il}),{\sigma^{2}}^{(t)}_{l}\bI_{m_{i}} \right)}$
\ENDFOR

\STATE 2. Sample the mixing proportions: $\bspi^{(t)}\sim \text{Dir}(\alpha_{1}+n^{(t)}_1,\ldots, \alpha_{K}+n^{(t)}_K)$ with $n^{(t)}_k = \sum_{i=1}^n z^{(t)}_{ik}$

\FOR{$k=1$ {\bfseries to} $K$}

\STATE 3. Sample the random-effects variance: 
${\xi^2_{k}}^{(t)} \sim IG \left( a_{0}+\frac{n}{2},  b_{0} + \frac{\sum_{i=1}^{n} {\bb_{ik}^{(t-1)}}^{T}\bb_{ik}^{(t-1)}}{2}\right)$

\STATE 4. Sample  the noise variance: \\
${\sigma^2_{k}}^{(t)} \sim IG \left(g_{0}+\frac{n_k}{2} , h_{0}+\frac{\sum_{i=1}^{n}{z_{ik}} \left( \by_{i}- \bS_{i}\bsbeta_{k}^{(t-1)}-\bS_{i}\bb_{ik}^{(t-1)} \right) ^{T}\left( \by_{i}- \bS_{i}\bsbeta_{k}^{(t-1)}-\bS_{i}\bb_{ik}^{(t-1)}\right)}{2}\right) $

\STATE 5. Sample the fixed-effects coefficient vector: 
$\bsbeta_{k}^{(t)}\sim \N(\boldsymbol{\nu}^{(t)}_{0},\bV^{(t)}_{0}) $ with\\
${\bV_{0}^{-1}}^{(t)}=\bsSigma_{0}^{-1}+\frac{1}{{\sigma^2_{k}}^{(t)}}\sum_{i=1}^{n}z^{(t)}_{ik}\bS_{i}^{T} \bS_{i}$, \\
$\boldsymbol{\nu}^{(t)}_{0}= \bV_{0}\left(\frac{1}{{\sigma^2_{k}}^{(t)}}\sum_{i=1}^{n}z^{(t)}_{ik}\bS_{i}^{T} \left( \by_{i}-\bS_{i}\bb_{ik}^{(t-1)}\right)-\bsSigma_{0}^{-1}\boldsymbol{\mu_{0}}\right)$ 
 
\FOR{$i=1$ {\bfseries to} $n$}
\STATE 6. Sample the random-effects coefficient vector: $\bb_{ik}^{(t)}\sim \N(\boldsymbol{\nu}^{(t)}_{1},\bV^{(t)}_{1})$ with \\
${\bV_{1}^{-1}}^{(t)}= \frac{1}{{\sigma^{2}_{k}}^{(t)}}\bS_{i}^{T}\bS_{i}+\frac{1}{{\xi_{k}^{2}}^{(t)}}$\\
${\boldsymbol{\nu}}^{(t)}_{1}= \bV_{1}\left(\frac{1}{{\sigma^{2}_{k}}^{(t)}}\bS_{i}^{T}(\by_{i}-\bS_{i}\bsbeta_{k}^{(t)})\right)$
\ENDFOR
\ENDFOR
\ENDFOR
\end{algorithmic}
\end{algorithm}

 \bigskip
 
 \subsection{Model-based surface clustering using the BMSSR} 
\label{ssec: MBC using BMSSR}

In addition to Bayesian density estimation, The BMSSR model can also be used for Bayesian model-based surface clustering so that to provide a partition of the data into $K$ clusters. 
Model-based clustering using the BMSSR model consists in assuming that the observed data $\{\bS_i,\by_i\}_{i=1}^n$ are  generated from a $K$ component mixture of spatial spline regressions with mixed-effects with parameter vector $\bsvPsi$. The mixture components can be  interpreted as clusters and hence each cluster can be associated with a mixture component. 
The problem of clustering therefore becomes the one of estimating the BMSSR parameters $\bsvPsi$. This is performed here by Gibbs sampling which provides a MAP estimator $\hat\bsvPsi_{\text{MAP}}$, which can be obtained by averaging the Gibbs posterior sample after removing some initial samples corresponding to a burn-in period. 
A partition of the data can then be obtained from the posterior memberships by applying the MAP rule, that is, by maximizing the posterior cluster probabilities (\ref{eq: postprob BMSSR}) to assign each observation to a cluster: 
\begin{eqnarray}
\hat{z}_i = \arg \max_{k=1}^K  \tau_{ik}(\hat\bsvPsi_{\text{MAP}})
\label{eq: MAP rule for clustering}
\end{eqnarray}where $\hat{z}_i$ represents the estimated cluster label for the $i$th observation.

\section{Application to simulated data and real data} 
\label{sec: application}
In this section we apply the two proposed Bayesian models\footnote{The corresponding algorithms (including the EM alternative) have been written in Matlab and are available upon request from the author.} on simulated data and real data. 
We first consider simulated surfaces to test the model in terms of surface approximation. Then, we apply it on a handwritten character recognition problem by considering real images from the MNIST data set \citep{LecunMnist} to test it in terms of surface approximation and clustering.

\subsection{Simulated surface approximation using the BSSR model}
We consider the bi-dimensional arbitrary function $\mu(\bx)=\dfrac{\sin(\sqrt{1+ x_1^2+x_2^2})}{\sqrt{1+ x_1^2+x_2^2}}$ and we attempt to approximate it from a sample of simulated noisy surfaces. 
We simulate a sample of 100 random surfaces $\by_i (i=1,\ldots,100)$ as follows. 
Each surface $\by_i$ is composed of $m_i = 21 \times 21$ observations generated  on a square domain $(x_1,x_2) \in [-10,10]\times[-10,10]$. To generate the surface $\by_i$, we first add random effects to the mean surface by computing $\bsmu_i(\bx) + \bb_i$ and then  $\by_i$ is simulated by adding a random error term, that is, $\by_i = \bsmu_i(\bx) + \bb_i + \be_i$ with $\bb_i \sim \N(\bO,0.1^2\bI_{m_i})$ and $\be_i \sim \N(\bO,0.1^2\bI_{m_i})$. 
Then, the sample of simulated surfaces $\bY=(\by_1,\ldots,\by_{100})$ is approximated by applying the BSSR model.

\begin{figure}[H]
\centering
\begin{subfigure}[b]{0.45\textwidth}
\includegraphics[width=\textwidth]{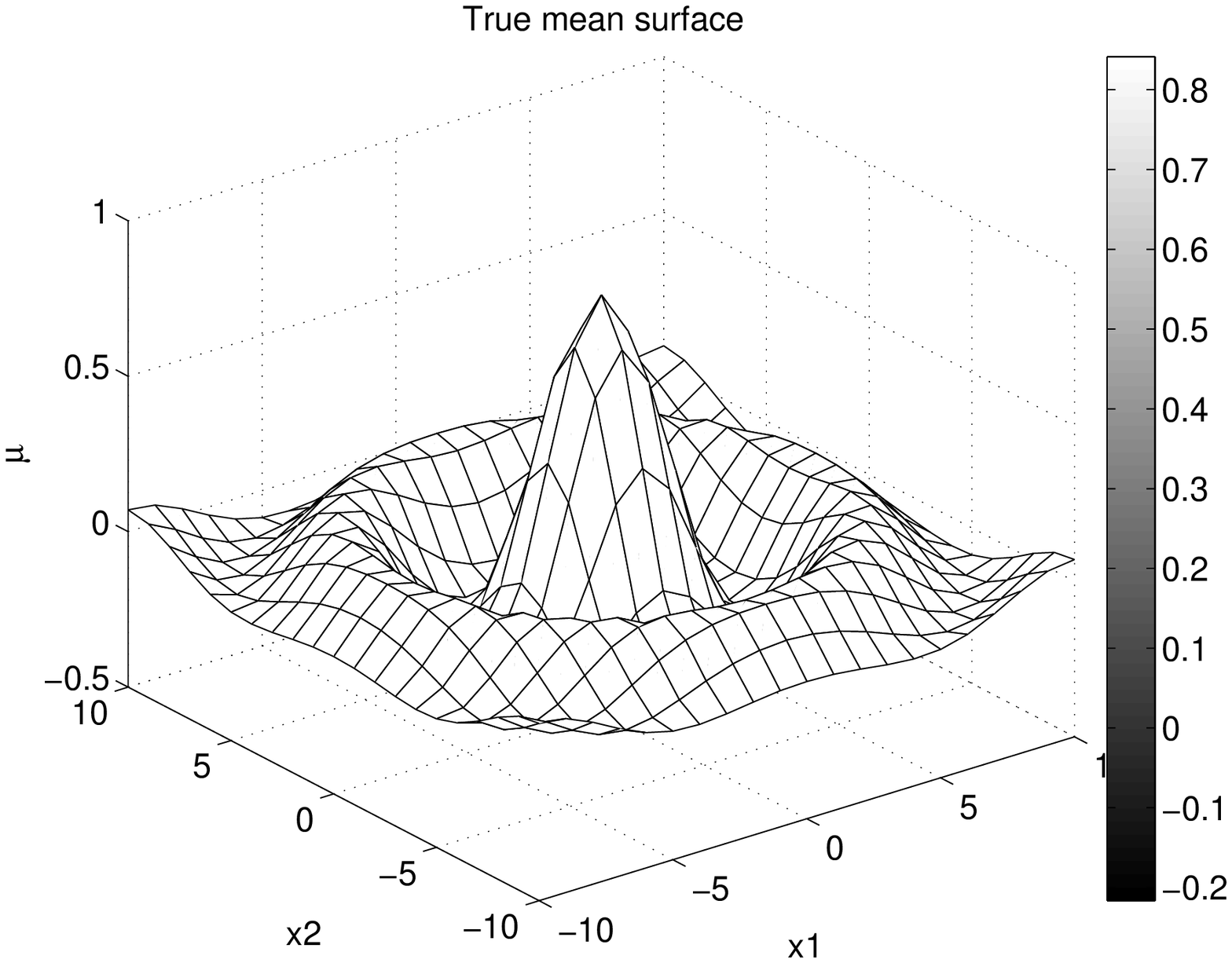}
\caption{}
\label{fig: True surface}
\end{subfigure}%
\begin{subfigure}[b]{0.45\textwidth}
\includegraphics[width=\textwidth]{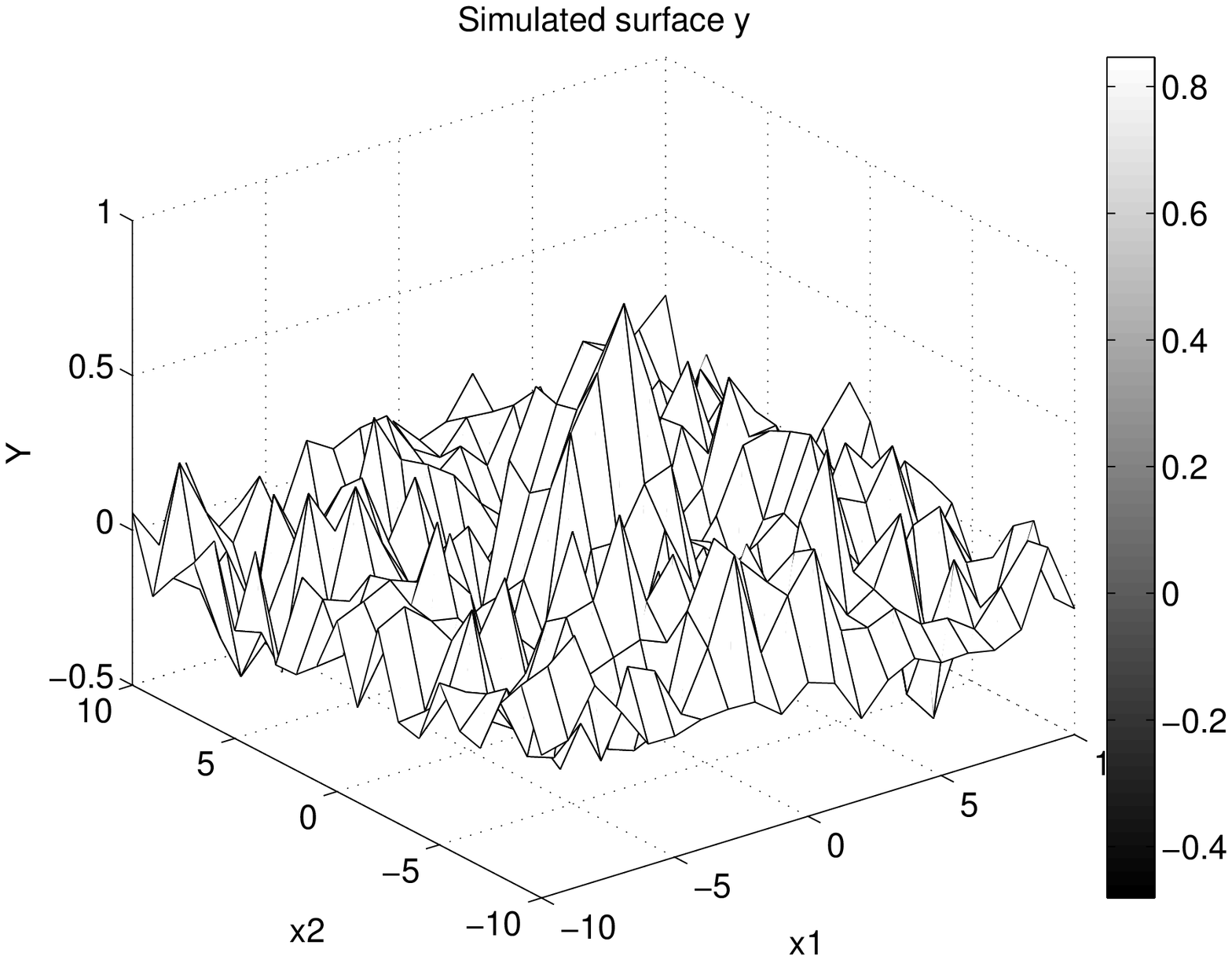}
\caption{}
\label{fig: simulated surface}
\end{subfigure} 
\\
\begin{subfigure}[b]{0.45\textwidth}
\includegraphics[width=\textwidth]{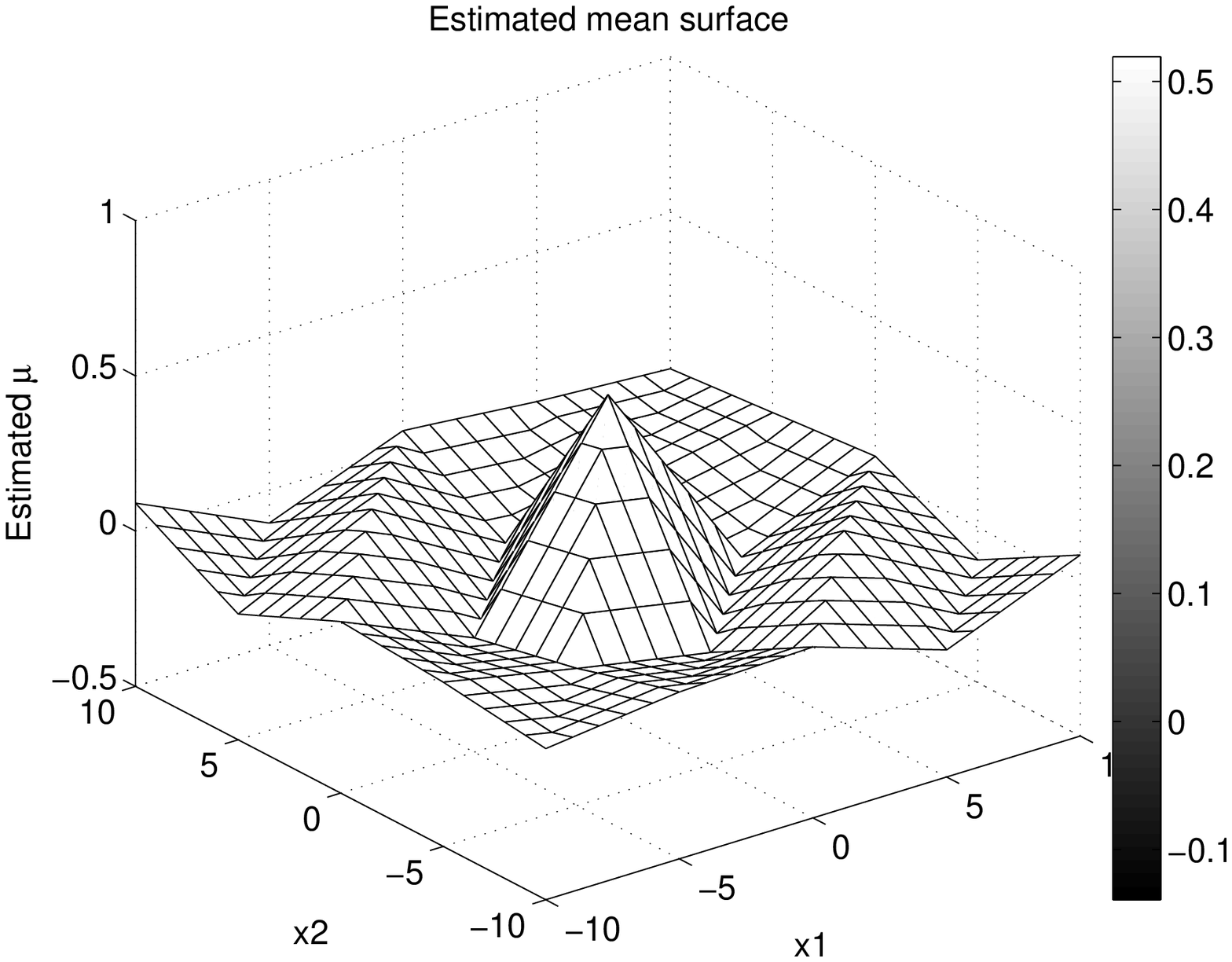}
\caption{}
\label{fig:regBSSR5x5}
\end{subfigure}
\begin{subfigure}[b]{0.45\textwidth}
\includegraphics[width=\textwidth]{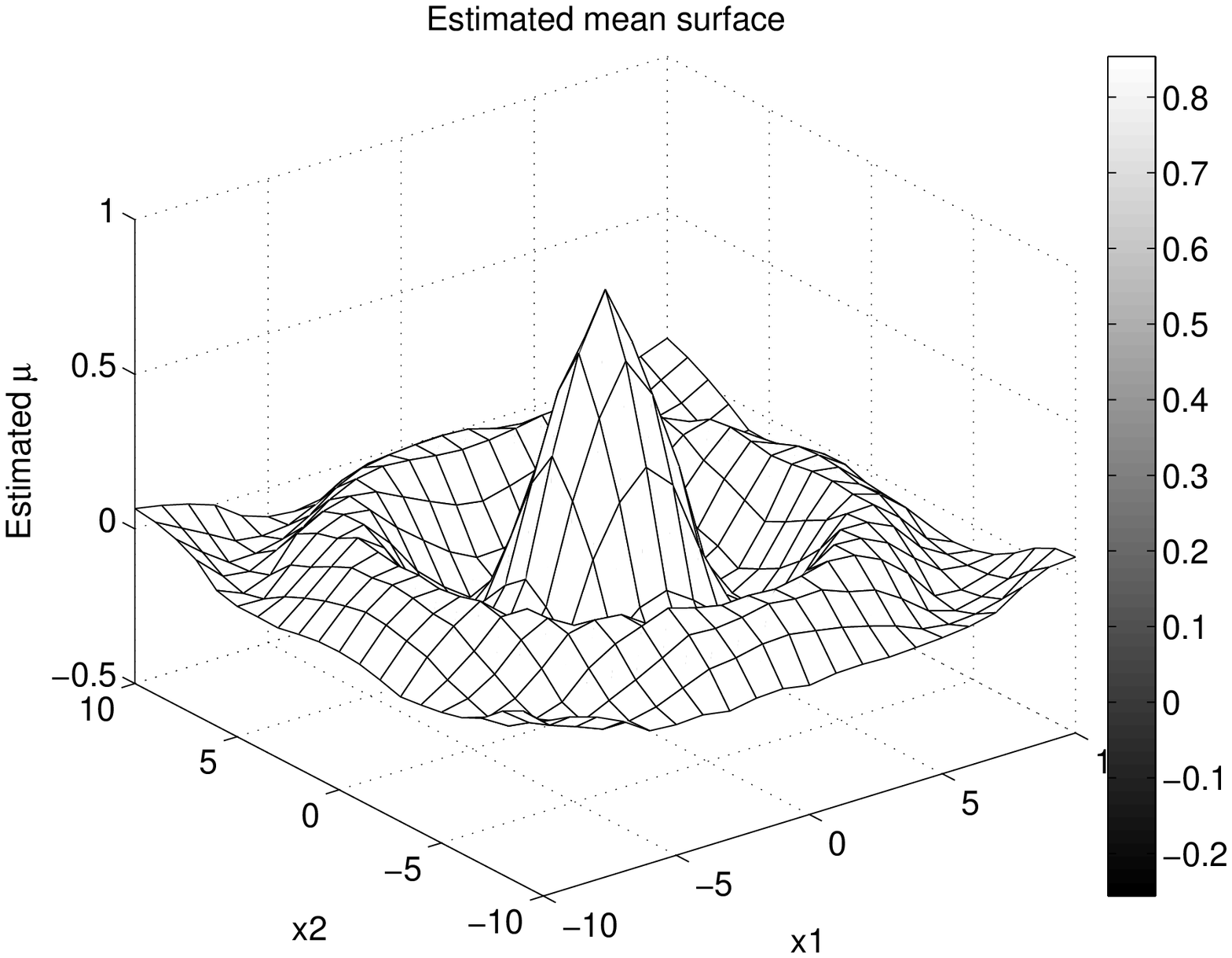}
\caption{}
\label{fig:regBSSR15x15}
\end{subfigure}%
\caption{True surface (a), an example of noisy surface from the simulated sample (b), A BSSR fit using $5\times 5$ NBFs (c) and $15\times 15$ NBFs (d).}
\label{fig:surfreg}
\end{figure}

Figure~\ref{fig: True surface} shows the actual mean function before the noise and the random effects are added, and Figure~\ref{fig: simulated surface} shows an example of  simulated surface. 
 We apply the BSSR model with $d=5\times5$ NBFs and $d=15\times15$ NBFs and show the obtained mean surface $\hat{\mu}(\bx) = \bS_i \hat{\bsbeta}$ fitted from the whole data set.\\
Figure~\ref{fig:regBSSR5x5} shows the fitted mean surface $\hat{\mu_1}$  with $d=5\times 5$ NBF basis, while Figure~\ref{fig:regBSSR15x15} shows its analogous with $d=15\times 15$ NBFs.
It can be seen that for the two cases, the approximated surface resembles the actual one. In particular, the second approximation, using a reasonable number of basis functions, is very close to the true surface.
This is confirmed by the value of the empirical sum of squared error between the true surface and the fitted one
$SSE = \sum^m_{j=1} (\mu_j(\bx)-\hat{\mu}_j(\bx))^2$ ($m=441$ here), which equal $0.0865$ in this case and which corresponds to a very reasonable fit.

\subsection{Handwritten digit clustering using the BMMSSR model}
\begin{figure}[H]
\centering
\includegraphics[width=0.6\linewidth]{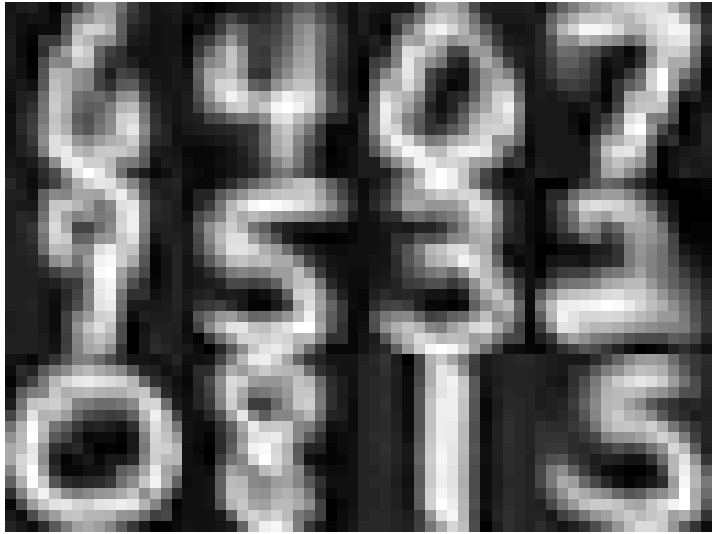}
\caption{Cluster means obtained by the proposed  BMSSR model for with $K=12$ components.}
\label{fig:B12C}
\end{figure}

 In this section we apply the BMSSR model 
on a subset of the ZIPcode data set \cite{hastieTibshiraniFreidman_book_2009}, which is issued from the MNIST data set \citep{LecunMnist}. The data set contains 9298 16 by 16 pixel gray scale images of Hindu-Arabic handwritten numerals distributed as in the following table~\ref{data:Zipcode}.
\bigskip \\ 
\begin{table}[H]
\centering
\begin{tabular} {|l|c|c|c|c|c|c|c|c|c|c|}
\hline digit & 1 & 2 & 3 & 4 & 5 & 6 & 7 & 8 & 9 & 0 \\ 
\hline training set & 1005 & 731 & 658 & 652 & 556 & 664 & 645 & 542 & 644 & 1194 \\ 
\hline testing set & 264 & 198 & 166 & 200 & 160 & 170 & 147 & 166 & 177 & 359 \\ 
\hline  
\end{tabular}
\caption{Zipcode data set digits distribution}
\label{data:Zipcode} 
\end{table}Each individual $\by_i$ contains $m_i=256$ observations $\by_i = (y_{i1},\dots,y_{i256})^T$ values in the range $[-1,1]$.
We run the Gibbs sampler given by algorithm \ref{algo: algo: Gibbs for BMSSR} with  a number of clusters $K = 8,\ldots,12$  on a subset of 1000 digits randomly chosen from the Zipcode testing set with the distribution given  in table~\ref{data:Zipcodesubset1},
\begin{table}[H]
\centering
\begin{tabular}{|l|c|c|c|c|c|c|c|c|c|c|}
\hline digit & 1 & 2 & 3 & 4 & 5 & 6 & 7 & 8 & 9 & 0 \\ 
\hline $K=8$ & 108 & 105 & 96 & 100 & 107 & 94 & 106 & 90 & 97 & 97 \\ 
\hline $K=9$ & 97  & 107 & 107 & 100 & 103 & 112 & 88 & 98 & 92 & 96  \\ 
\hline $K=10$ & 97 & 90 & 100 & 98 & 107 & 107 & 102 & 107 & 97 & 98  \\ 
\hline $K=11$ & 105 & 104 & 99 & 96 & 95 & 106 & 101 & 93 & 107 & 94 \\ 
\hline  $K=12$ & 111 & 96 & 96 & 105 & 108 & 96 & 97 & 99 & 97 & 98 \\ 
\hline   
\end{tabular}
\caption{Repartion of used Zipcode random  subsets of 1000 digits}
\label{data:Zipcodesubset1} 
\end{table}We used $d=8 \times 8$ NBFs, which corresponds to the quarter of the resolution of the images in the Zipcode data set.  
We performed five runs of 
of the algorithm, each for a different model: $K=8,\ldots,12$. The corresponding mean Adjusted Rand Index (ARI) values are given in  Table \ref{tab: BMSSR ARI for ZIPcode}. The model with $12$ clusters has the highest ARI value.  
\begin{table}[H]
\centering
\begin{tabular}{|l|c|c|c|c|c|}
\hline $K$& 8 & 9 & 10 & 11 & 12 \\  
\hline ARI &  0.4848 & 0.4694 &  0.4445 &  0.5139 & 0.5238 \\  
\hline
\end{tabular}
\caption{ARI for the BMSSR model for a number of clusters $ K=8,\ldots,12.$}
\label{tab: BMSSR ARI for ZIPcode} 
\end{table} 
Figure~\ref{fig:B12C} 
shows the cluster means for $K=12$ obtained by the proposed Baysian model (BMSSR).
It clearly shows that the model is able to recover the ten digits as well as subgroups of the digit $0$ and the digit $5$.

 \section{Conclusion and future work}
 \label{sec: conclusion}
We presented a probabilistic Bayesian model for homogeneous spatial data based on spatial spline regression with mixed-effects (BSSR). The model is able to accommodate individual with both fixed and random effect variability. Then, motivated by a model-based surface clustering perspective, we introduced the Bayesian mixture of spatial spline regressions with mixed-effects (BMSSR) for spatial functional data dispersed into groups. 
We derived Gibbs samplers to infer the models. Application on simulated surfaces illustrates the surface approximation using the BSSR model. Then, application on real data in a handwritten digit recognition framework shows the potential benefit of the proposed BMSSR model for practical applications on surface clustering.
\\
The  BMSSR can be extended to be used for supervised surface classification. This can be performed without difficulty by modeling each class by a BMSSR model and then applying the Bayes rule to assign a new observation to the class corresponding to the highest posterior probability.
\\
A future work will therefore consist in conducting additional experiments on real data clustering and discrimination as well as model selection using information criteria such as BIC and ICL. \\ Then, another interesting perspective is to derive a Bayesian non-parametric model by relying of Dirichlet Process mixture models where the number of mixture components can be directly inferred from the data.

\appendix

\bibliographystyle{apalike}

\bibliography{bayes-SSRM}  
\end{document}